\documentclass[draftcls,onecolumn,12pt]{IEEEtran}

\usepackage{algorithmic}
\usepackage{algorithm}
\usepackage{indentfirst}
\usepackage{cite}
\usepackage{amsmath}
\usepackage{amsfonts}
\usepackage{amssymb}
\usepackage{times}
\usepackage{subfigure}
\usepackage[english]{babel}
\usepackage{stfloats}
\usepackage{bbm}
\usepackage{dsfont}
\usepackage[dvips]{graphicx}
\usepackage{array}
\usepackage[normalem]{ulem}
\usepackage{epsf}
\usepackage{psfrag}
\usepackage{comment}
\usepackage{color}
\usepackage{xcolor}
\usepackage{tikz}
\usetikzlibrary{automata,arrows,positioning,calc}
\usepackage{epstopdf}

\begin{document}

\newcommand{\figw}{0.99\columnwidth}
\newcommand{\figww}{0.7\columnwidth}
\newcommand{\figwww}{0.6\columnwidth}
\newcommand{\figwwww}{0.49\columnwidth}
\newcommand{\figsmall}{0.32\columnwidth}

\title{On the Effect of Fronthaul Latency on ARQ in C-RAN Systems}
\author{\large Qing Han$^*$, Chenxi Wang$^{\dagger}$, Marco Levorato$^*$ and Osvaldo Simeone$^{\ddagger}$\\
\normalsize $*$ The Donald Bren School of Information and Computer Science, UC Irvine, CA, US\\
\normalsize $\dagger$ The Henry Samueli School of Engineering, UC Irvine, CA, US\\
\normalsize $\ddagger$ Dept.\ of Electrical Engineering, New Jersey Institute of Technology, Newark, USA.}
\date{}

\maketitle

\begin{abstract}
In the Cloud Radio Access Network (C-RAN) architecture, a Control Unit (CU) implements the baseband processing functionalities of a cluster of Base Stations (BSs), which are connected to it through a fronthaul network. This architecture enables centralized processing at the CU, and hence the implementation of enhanced interference mitigation strategies, but it also entails an increased decoding latency due to the transport on the fronthaul network. The fronthaul latency may offset the benefits of centralized processing when considering the performance of protocols at layer 2 and above. This letter studies the impact of fronthaul latency on the performance of standard Automatic Retransmission reQuest (ARQ) protocols, namely Stop and Wait, Go-Back-N and Selective Repeat. The performance of the C-RAN architecture in terms of throughput and efficiency is compared to the that of a conventional cellular system with local processing, as well as with that of a proposed hybrid C-RAN system in which BSs can perform decoding. The dynamics of the system are modeled as a multi-dimensional Markov process that includes sub-chains to capture the temporal correlation of interference and channel gains. Numerical results yield insights into the impact of system parameters such as fronthaul latency and signal-to-interference ratio on different ARQ protocols.
\end{abstract}

\section{Introduction}

Cloud Radio Access Network (C-RAN) is a cloud
computing-based radio access network architecture in which, unlike traditional cellular systems,
the protocol stack of a cluster of base stations (BSs) is virtualized by means of a Control Unit (CU), to which the BSs are connected via a fronthaul network. This architecture enables centralized signal processing at the CU, hence allowing the implementation of effective interference management strategies, which may dramatically improve the physical-layer capacity of the system \cite{chinamobile} -\cite{front}. However, the transmission and queueing delays on the fronthaul network generally limit the gains that can be achieved when considering the performance at higher layers, particularly at Layer 2 \cite{jm} -\cite{ngmn}.

Fronthaul latency affects most notably the operations of Automatic Retransmission reQuest (ARQ) protocols in the uplink, which rely on the transmission of acknowledgement (ACK) or negative-acknowledgement (NAK) messages to the mobile terminals (MTs), indicating respectively correct or incorrect packet decoding. In fact, due to fronthaul transmission and processing, these messages incur a delay that may significantly reduce throughput and efficiency of retransmission mechanisms. For example, in Long-Term Evolution (LTE) systems, a latency larger than 3-8 ms disrupts the normal working of the Hybrid ARQ (HARQ) protocol \cite{5g}.

In~\cite{5g,HARQ,Khalili}, in order to mitigate the effect of fronthaul latency on the performance of HARQ protocols, it is proposed that the BSs make local decisions regarding the expected outcome of decoding at the CU without waiting for the actual feedback from the CU. The local decisions are made based on the knowledge at the BSs of the modulation and coding scheme used for uplink transmission and on an estimate of the channel state information. Note that this requires the BSs to implement enough baseband functionalities to be able to perform resource demapping, e.g., synchronization and FFT, as well as channel estimation, hence deviating from the conventional C-RAN paradigm \cite{5g}\cite{5g1}. The drawback of this approach is that, especially for small coding blocks, the local decisions taken at the BSs may be incorrect, causing inefficiencies in the use of the spectral resources and delays \cite{Khalili}.

The analysis in \cite{HARQ,Khalili} is oblivious to the magnitude of the fronthaul latency, as it is implicitly based on the assumption of instantaneous ACK/NAK feedback from the BSs to the MTs. In this letter, instead, we are interested in evaluating the performance of ARQ protocols as a function of the fronthaul latency. Fronthaul latency translates into feedback delays, whose impact on the performance of ARQ, unlike in the analysis in \cite{HARQ, Khalili}, depends on the specific implemented ARQ protocol. This is because different ARQ protocols react in distinct ways to feedback delays. In particular, we consider the throughput and efficiency of standard ARQ protocols, namely Stop and Wait (SW), Go-Back-N (GBN) and Selective Repeat (SR) (see, e.g., \cite{arq}), as a function of the fronthaul delay. Furthermore, motivated by the key role that interference mitigation plays in the evaluation of the performance benefits of C-RAN, unlike prior work, we explicitly consider the impact of interference. Specifically, the dynamics  of the interference process, along with that of the channel gains, are modelled by means of Markov processes. The analysis encompasses the standard C-RAN implementation with centralized baseband processing, as well as conventional cellular systems and an hybrid approach that is akin to the solution studied in \cite{5g,HARQ,Khalili}. In the hybrid approach, a BS first attempts local decoding, and processing at the CU is invoked only if this decoding attempt is unsuccessful. It is noted that, as for the approach in  \cite{5g,HARQ,Khalili}, this hybrid solution requires the BSs to be endowed with baseband processing functionalities.

The rest of this paper is organized as follows. Section~\ref{system} describes the system model and the ARQ protocols
considered in this paper. Section~\ref{model} presents the analysis based on Markov modelling that is used to assess throughput and efficiency.
In Section~\ref{evaluation}, numerical results are shown to illustrate the performance of the system. Section~\ref{conclusion} concludes the paper.

\section{System Model}\label{system}

\begin{figure}[h]
\centering
\includegraphics[width=3in]{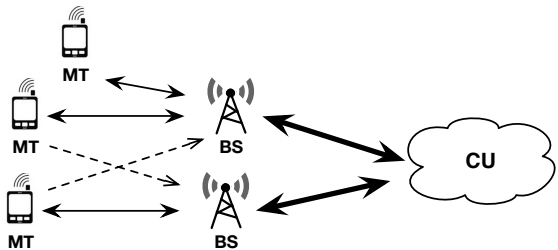}
\caption {Illustration of a C-RAN. Bold lines represent fronthaul links, while solid lines represent wireless links and dashed lines represent
wireless interference.}
\label{fig:system}
\vspace{-2mm}
\end{figure}

The uplink of the cellular system depicted in Fig.~\ref{fig:system} is considered, which is composed of a cluster of BSs connected to a CU through fronthaul links. Slotted time is assumed, where the transmission of one packet fits the duration of one time slot. The MTs store packets to be transmitted in a finite First-In First-Out buffer of size $B_{\rm max}$ packets. Packet arrival in the buffer is modeled as an independent identically distributed (i.i.d.) process, where the probability that a new packet arrives in one slot is equal to $\alpha$ ($ 0 \leq \alpha \leq 1$). The MTs implement ARQ to improve packet reliability, where failed packets are retransmitted until successfully decoded. We specifically study the standard ARQ protocols SW, GBN and SR.

We consider three different system implementations, namely: (\emph{i})
\emph{Conventional cellular system}: An assigned BS performs decoding of the uplink signal of any given MT; (\emph{ii}) \emph{C-RAN}: The CU performs decoding based on the signals received by all BSs on the fronthaul links; (\emph{iii})
\emph{Hybrid C-RAN}: An assigned BS first attempts decoding and, if it fails, the received signal is forwarded to the CU,
which attempts decoding based on all the BSs' signals. As mentioned, the latter solution requires the implementation  of an alternative functional split between BS and CU, as compared to conventional C-RANs, in which the BS implements baseband functionalities \cite{5g, 5g1}.

We assume that the propagation time between MT and BS is negligible, so that
packets transmitted in slot $k{=}0,1,\ldots$ are acknowledged by means of ACK/NAK messages at the end of the same slot, if decoding takes place at the BS as in systems (\emph{i}) and (\emph{iii}). Instead, to capture the delay associated with two-way transmission over the fronthaul network, we assume that ACK/NAK messages relative to decoding at the CU are received at the MTs after $\Delta \geq 1$ time slots in systems (\emph{ii}) and (\emph{iii}).

A detailed description and discussion of the signal processing techniques used in a C-RAN architecture of BSs and CU
can be found in, e.g.,~\cite{jm, front}. Here, focusing on the performance at layer 2, we abstract the physical layer
by adopting a simple threshold model on the signal to noise ratio. Specifically, a packet transmitted by an MT in
a slot $k$ is successfully decoded at an assigned BS in architectures (\emph{i}) and (\emph{iii}) if the signal-to-interference-plus-noise ratio (SINR) at the BS satisfies the condition
\begin{equation}\label{SINR}
{\rm SINR}^{BS}(k) = \frac{Ph(k)}{I(k)+\sigma^2} > \gamma,
\end{equation}
where $\gamma$ is the decoding threshold; $\sigma^2$ is the variance of the noise; $h(k)$
is the channel gain between the given MT and the assigned BS in slot $k$; $P$ is the uplink transmitted power; and the term $I(k)$ captures the overall interference incurred
by the packet sent by the MT in slot $k$. Instead, to account for decoding at the CU in architectures  (\emph{ii}) and (\emph{iii}), we assume that the CU can effectively remove the interference based on the signals received from multiple BSs. As a result, a packet is decoded at the CU if the signal-to-noise ratio (SNR) at the CU satisfies
\begin{equation}\label{SINR2}
{\rm SNR}^{CU}(k) = \frac{Ph(k)}{\sigma^2+\sigma_q^2(k)} > \gamma,
\end{equation}
where the additional noise term $\sigma^2_{q}(k)$ 
captures the effect of quantization noise, see, e.g., \cite{front}. The latter is generally an exponentially decreasing function of the fronthaul capacity and is proportional to the received power \cite{front}. Here, in order to focus on the effect of the fronthaul latency, we assume this term to be small enough to be negligible as compared to the channel noise. The analysis could be easily extended to a scenario in which this assumption is violated.

The temporal evolution of the interference process $I(k)$ is modeled as an on-off process
\begin{equation}\label{int}
{I(k)} = I\Psi(k),
\end{equation}
where $\Psi(k){\in}\{0,1\}$
is a binary Markov chain. This simple model captures the correlation in the interference process that results from the packet arrival
process at the interfering nodes, as well as from the temporal correlation of the interferers channel gains. We note that the model can be extended to include multiple levels of interference
and that the transition probabilities, to be discussed below, are generally a function the transmission patterns of the interfering nodes.

To measure the performance of the network, we use the following metrics: (\emph{a}) \emph{Throughput}, i.e., the average number of packets successfully decoded per slot; (\emph{b}) \emph{Efficiency}, i.e., the average number of packets successfully decoded per slot divided by the fraction of slots in which the MT transmits. We observe that the efficiency directly reflects the energy requirements at the MT since a low efficiency entails the need for more packet transmissions.

We now briefly recall the three ARQ protocols under study. For a more detailed discussion, please see \cite{arq}.

\subsubsection{Stop-and-Wait}
With SW, if the buffer is non-empty, the oldest packet is transmitted by the MT in the uplink and the
MT waits idle to receive an ACK/NAK message. Therefore, in the conventional cellular architecture (\emph{i}) and in the hybrid C-RAN system (\emph{iii}), if the BS successfully
decodes the packet, an ACK is received in the same slot, and the MT can transmit the next packet in the queue in the following slot. Instead, in architecture (\emph{i}), if the BS fails, the packet is retransmitted in the next slot, while, in system (\emph{iii}), the MT waits $\Delta$ slots and then transmits a new packet or retransmit the same packet upon reception of the ACK/NACK relative to decoding at the CU. Finally, in the C-RAN system (\emph{ii}), the MT always waits $\Delta$ slots for feedback from the CU, and transmits the same or the next packet depending on whether a NAK or ACK message is received.

\subsubsection{Go-Back-N}
With GBN, the MT keeps a list, or window, of at most $W$ outstanding packets that have been transmitted but whose ACK/NAK messages have not been received or processed yet at the MT. If the window contains less than $W$ packets, a new packet is transmitted from the queue if there is any. When the window is full, the MT checks for the ACK/NAK messages that have been received for the packets in the window. If there are no ACK/NAK messages, the MT stays idle. If  there are, packets are considered starting from the oldest in the window. For this packet, if an ACK has been received, the window slides to the following packet, which is considered next; instead, if a NAK has been received for the packet, the MT retransmits this and all the following packets in the current window. Note that GBN is meant to avoid the throughput loss associated with idle transmit periods caused in SW by delayed feedback, and is hence relevant, in our context, only for the C-RAN and hybrid C-RAN architectures. This is because, in these systems, an MT implementing SW must wait for feedback from the CU, whereas GBN allows the transmission of new packets, if the window is not full, even before reception of ACK/NAK messages for the last transmitted packet.

\subsubsection{Selective-Repeat}
With SR, the MT operates in the same was as for GBN with the caveat that, when the window is full and ACK/NAK messages are considered for the packets in the window, only the failed packet are retransmitted. As for GBN, hence, SR is relevant only for the C-RAN and hybrid C-RAN architectures.

\section{Analysis Based On Markov Modeling}
\label{model}
In this section, we evaluate throughput and efficiency of the ARQ protocols discussed above by focusing on the performance of an individual MT. The dynamics of the system are modeled as a Markov chain ${\bf S}{=}(S(1),S(2),\ldots)$ with
a finite state space $\mathcal{S}$. The transmission process of the MT, along with the channel $h(k)$
and the interference process $\Psi(k)$ are sub-chains that together compose the overall chain. The state of the system chain in slot $k$ is denoted as $S(k){=}(h(k),\Psi(k),\Phi(k))$,
where $\Phi(k)$ tracks the protocol state, e.g., the state of the queue at the MT and also of the retransmission window for the GBN and SR protocols.
The transition probability $\mathit{p}(s^{\prime}|s){=}\mathit{P}(S(k{+}1){=}s^{\prime}|S(k){=}s)$, for any two states $s=(h,\Psi,\Phi)$, $s^{\prime}=(h^{\prime},\Psi^{\prime}, \Phi^{\prime})\in\mathcal{S}$ can be factored as
\begin{equation}\label{factors}
\mathit{p}(s^{\prime}|s){=}\mathit{p}_h(h^{\prime}|h)\mathit{p}_{\Psi}(\psi^{\prime}|\psi)\mathit{p}_{\Phi}(\phi^{\prime}|h,\psi,\phi),
\end{equation}
where we have
\begin{align}
\mathit{p}_h(h^{\prime}|h) &\!\!=\!\! \mathit{P}(h(k{+}1){=}h^{\prime}|h(k){=}h),\\
\mathit{p}_{\Psi}(\psi^{\prime}|\psi) &\!\!= \!\!\mathit{P}(\Psi(k{+}1){=}\psi^{\prime}|\Psi(k){=}\psi),\\
\mathit{p}_{\Phi}({\phi}^{\prime}|h,\psi,\phi) &\!\!=\!\! \mathit{P}(\Phi(k{+}1){=}\phi^{\prime}{|}h(k){=}h,\Psi(k){=}\psi,\Phi(k){=}\phi).
\end{align}
In the following, we briefly elaborate on the dynamics of channel, interference and protocol, as defined by the factors in Eq.~(\ref{factors}).


\subsection{Channel Sub-Chain}

We adopt a correlated Rayleigh fading model with correlation coefficient $\rho$ across two successive slots \cite{rayleigh} and unitary power. In order to make the model manageable within the Markov chain framework at hand, we use the quantized model detailed in \cite{rayleigh}, in which the channel gain range is partitioned into $Q$ states with equal steady state probabilities $\pi_s{=}1/Q$, $s{=} 1, 2, ..., Q$. The corresponding transition probabilities $\mathit{p}_h(h^{\prime}|h)$ can be calculated as detailed in~\cite{rayleigh}.

\begin{figure}[!t]
\begin{center}
\begin{tikzpicture}[->, >=stealth', auto, semithick, node distance=3.5cm]
\tikzstyle{every state}=[fill=white,draw=black,thick,text=black,scale=0.65]
\node[state]    (0)                    {$0$};
\node[state]    (1)[right of=0]   {$1$};
\node[state, draw=none]  (2)[right of=1] {$\cdots$};
\node[draw=none] (3)[above right =1cm of 0]{};
\node[draw=none] (4)[below right =1cm of 0]{};
\node[state]    (Q)[right of=2]  {$Q$};

\path
(0) edge[loop left,below]              node{\footnotesize{$\mathit{p}_{h}(0|0)$}}   (0)
     edge[bend left,above]   node{\footnotesize{$\mathit{p}_{h}(1|0)$}}  (1)
     edge[bend left,left]   node{\footnotesize{$\mathit{p}_{h}(i|0)$}}  (3)
(1) edge[loop above]            node{\footnotesize{$\mathit{p}_{h}(1|1)$}}   (1)
     edge[bend left,below]   node{\footnotesize{$\mathit{p}_{h}(0|1)$}}  (0)
     edge[bend left,above]   node{\footnotesize{$\mathit{p}_{h}(i|1)$}}  (2)
(2) edge[bend left,below]   node{\footnotesize{$\mathit{p}_{h}(1|j)$}}  (1)
     edge[bend left,above]   node{\footnotesize{$\mathit{p}_{h}(Q|j)$}}  (Q)
(4) edge[bend left,left]     node{\footnotesize{$\mathit{p}_{h}(0|j)$}}  (0)
(Q) edge[loop above]            node{\footnotesize{$\mathit{p}_{h}(Q|Q)$}}   (Q)
     edge[bend left,below]   node{\footnotesize{$\mathit{p}_{h}(i|Q)$}}  (2);
\end{tikzpicture}
\setlength{\unitlength}{1mm}
\begin{picture}(80,7)
\linethickness{1pt}
\put(1,5){\vector(1,0){78}}
\put(1,2){\small{`bad'}}
\put(71,2){\small{`good'}}
\multiput(21,2)(0,2){3}{\line(0,1){1}}
\put(19,0){\small{$h_{0}$}}
\multiput(45,2)(0,2){3}{\line(0,1){1}}
\put(42,0){\small{$h_{1}$}}
\end{picture}
\end{center}
\caption {Channel sub-chain modelling a time-correlated Rayleigh fading channel as in \cite{rayleigh}. }
\label{Nmarkov}
\end{figure}
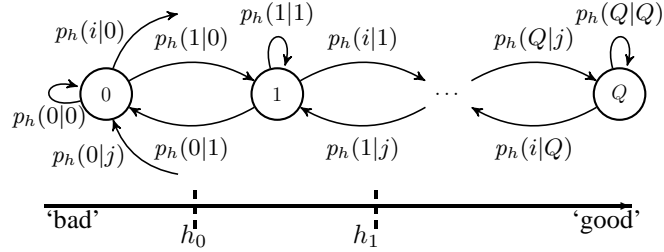

The threshold model in Eq.~(\ref{SINR}) and (\ref{SINR2}) determines a map between the channel and interference states, on the one hand, and the outcome
of packet decoding at the BS and CU, on the other. To clarify this map, we define the thresholds $h_{0}{=}\lfloor\frac{\gamma \sigma^2}{P}\rfloor$ and $h_{1}{=}\lfloor\frac{\gamma (\sigma^2+I)}{P}\rfloor$. If the channel state satisfies $h(k){=}h{>}h_0$, then the packet transmitted in
slot $k$ is decoded at the BS only if $\psi(k){=}0$, that is, in the absence of interference, whereas decoding fails if $\psi(k){=}1$. Instead,
due to interference cancellation at the CU, packet decoding at the CU is successful if $h(k){=}h{>}h_0$ irrespective
of the value of $\psi(k)$. If $h(k){=}h{>}h_1$, then the BS successfully decodes the packet transmitted by the MT
in slot $k$. Fig.~\ref{Nmarkov} illustrates the channel sub-chain by emphasizing the thresholds $h_{0}$ and $h_{1}$.

\subsection{Interference Sub-Chain}

The interference process $\Psi(k){\in}\{0,1\}$ is modeled as a binary Markov chain, where $\Psi(k){=}1$
corresponds to a slot with interference. We parametrize the chain $\Psi(k)$ through the
interference rate $E$ and the burstiness $B$, where $E$ is the fraction of time spent by the chain in state $1$,
and $B$ is the average length of sequences of state $1$. With this parametrization, the dynamics of $\Psi$ are defined by the transition probabilities $[\mathit{p}_{\Psi}(j|i)]_{i,j{=}\{0,1\}}$, where \cite{badia}
\begin{equation}
{E=\frac{\mathit{p}_{\Psi}(1|0)}{\mathit{p}_{\Psi}(1|0)+\mathit{p}_{\Psi}(0|1)} },~~~ {B=\frac{1}{\mathit{p}_{\Psi}(0|1)}}.
\end{equation}



\subsection{Protocol Sub-Chain}

The protocol sub-chain models the temporal evolution of the transmission process of the MT, as a function of the channel
and interference states, according to the rules prescribed by the implemented ARQ protocol and by the system architecture, namely conventional cellular system, C-RAN or hybrid C-RAN. In general, the protocol chain tracks the number of packets in the buffer of the MT and the internal state of the ARQ mechanism, such as the content of the window for the GBN and SR protocols.

As an example, consider SW and hybrid C-RAN. In this case, the protocol sub-chain tracks the number of packets in the buffer and the time remaining before the next transmission. To elaborate, we define the protocol state as $\Phi(k)=(B(k),C(k))$, where $B(k){\in}\{0,\ldots,B_{\rm max}\}$ is the number of packets in the MT's buffer at the beginning of slot $k$, and $C(k){\in}\{0,1,\ldots,\Delta\}$ is the number of slots left before the next transmission. If $B(k){>}0$ and $C(k){=}0$, then the MT transmits a packet in slot $k$. If the packet is decoded at the BS, then one packet is removed from the buffer and $B(k)$ is updated by accounting for a possible packet arrival. If decoding at the BS fails, then $C(k)$ is set to $\Delta$, indicating that the MT waits to receive feedback regarding decoding at the CU, and $C(k)$ is sequentially decreased in the following slots until $0$ is reached and a new attempt is performed.

A detailed description of the protocol chain for the other configurations is not included here due to space constraints, but it follows from the same considerations outlined above. For example, a related protocol chain tracking the state of SR ARQ can be found in~\cite{badia}.

\subsection{Performance Metrics}

By definition, the throughput is the average number of packets successfully decoded per slot, which can be computed as
$\mathcal{T}{=}\sum_{s{\in}\mathcal{S}} \pi(s) c_{\rm thr}(s)$,
where $\pi(s)$ is the steady state probability of state $s{\in}\mathcal{S}$, namely
\begin{equation}
\pi(s){=}\lim_{k\rightarrow\infty} \mathit{P}(S(k){=}s|S(0){=}s_0), \label{eq:limit}
\end{equation}
and $c_{\rm thr}(s)$ is equal to the number of packets delivered in state $s$. Note that $\pi(s)$ is equal to
the fraction of time spent by the system in state $s$ and that the limit in (\ref{eq:limit}) exists under mild assumptions on the ergodicity of the system Markov chain. The efficiency is the average number of packets successfully decoded per slot divided by the fraction of slots in which the MT transmits, which is given by
\begin{equation}
\mathcal{E}{=}\sum_{s{\in}\mathcal{S}} \pi(s) c_{\rm thr}(s)/\sum_{s{\in}\mathcal{S}} \pi(s) c_{\rm tx}(s),
\end{equation}
where $c_{\rm tx}(s)$ is equal to the number of packets transmitted in state $s$.

\section{Numerical Results}\label{evaluation}

In this section, we present numerical results to assess the performance of the ARQ protocols SW, GBN and SR for the conventional cellular system and for the C-RAN and hybrid C-RAN architectures discussed above. The system parameters are set as $\alpha{=}0.5$, $B_{\rm max}{=}1$, $P{=}30$ dB, $\sigma^2{=}1$ and $\gamma{=}10$~dB; the channel is characterized by correlation $\rho{=}0.3$; the
parameters of the interference chain are $E{=}0.6$ and $B{=}6$; and, unless indicated otherwise, the delay introduced by the fronthaul
network is set to $\Delta{=}5$~slots.  For the GBN and SR protocols, in the C-RAN and hybrid C-RAN systems, the transmission window is set to be equal the maximum delay, that is, $W{=}\Delta$. This choice is dictated by the observation that selecting $W{<}\Delta$ would induce idle slots when the packet is not decoded correctly at the BS, whereas the choice $W{>}\Delta$
would unnecessarily increase latency.

\begin{figure}[!t]
	\centering
	\includegraphics[width=\figwww]{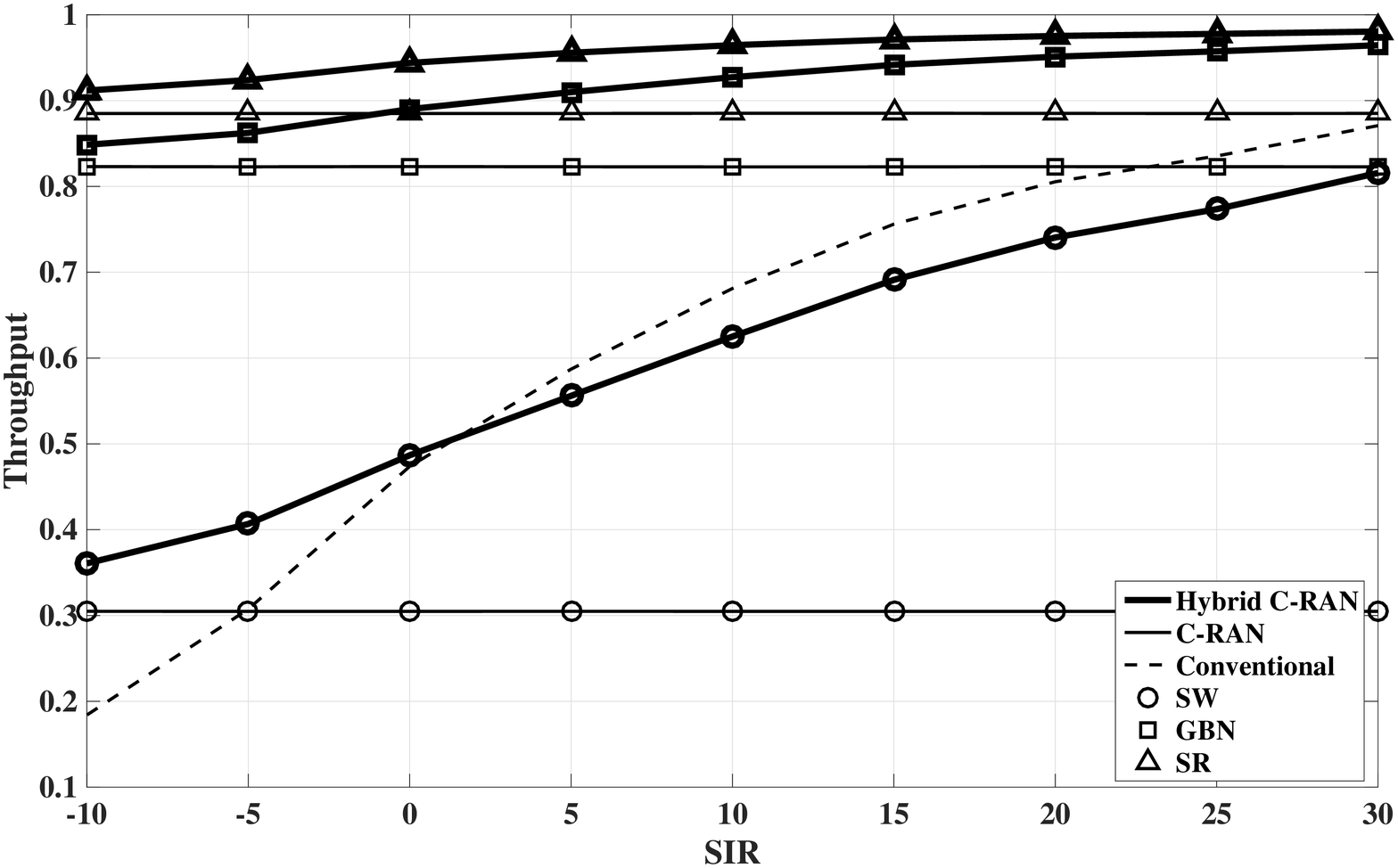}
	\caption{Throughput as a function of the average SIR ($\Delta{=}2$ slots, $\rho{=}0.3$, $\gamma{=}10$ dB, $\alpha{=}0.5$, $B_{\rm max}{=}1$, $W{=}5$, $P{=}30$ dB, $E{=}0.6$ and $B{=}6$).\vspace{-8mm}}
	\label{thrSIR}
\end{figure}

\begin{figure}[!t]
	\centering
	\includegraphics[width=\figwww]{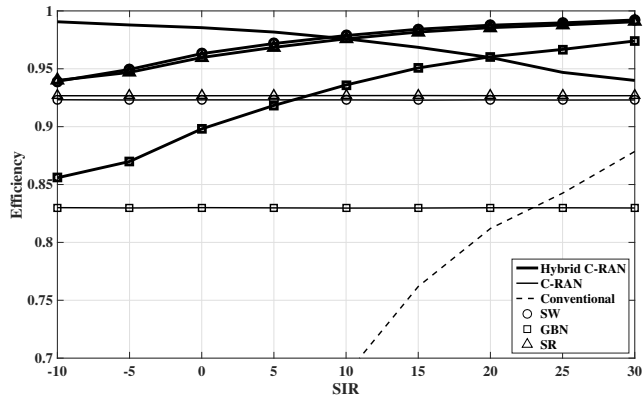}
	\caption{Efficiency as a function of the average SIR ($\Delta{=}2$ slots, $\rho{=}0.3$, $\gamma{=}10$ dB, $\alpha{=}0.5$, $B_{\rm max}{=}1$, $W{=}5$, $P{=}30$ dB, $E{=}0.6$ and $B{=}6$).\vspace{-8mm}}
	\label{effSIR}
\end{figure}

Fig.~\ref{thrSIR} and Fig.~\ref{effSIR} show the throughput and the efficiency, respectively, achieved by the conventional cellular system, the C-RAN and the hybrid C-RAN systems for the different ARQ protocols as a function of the $SIR = P/I$. As the SIR increases, the throughput and efficiency improve due to the lower error rate for decoding at the BS, which affect both the conventional and the hybrid C-RAN systems, but not the C-RAN architecture. Moreover, the performance of the hybrid C-RAN system converges to that of the conventional and C-RAN systems in the high and low SIR regions, respectively. In fact, if the SIR is high, decoding at the BS does not incur a significant performance degradation due to interference and hence decoding at the CU does not offer gains. As the SIR decreases, instead, the decoding probability at the BS becomes smaller, and the packet is always forwarded to the CU for decoding. Also, by construction, the hybrid C-RAN always outperforms the C-RAN system in both throughput and efficiency.

Considering now the impact of different ARQ protocols, if the basic SW is implemented, the fronthaul latency affects significantly the throughput, and the conventional approach offers the best throughput even at moderately low SIR values. For $\Delta = 5$ time slots, the impact of the delay on GBN is limited, and both C-RAN and hybrid C-RAN offer performance superior to that of the conventional system. Finally, the SR retransmission mechanism is the most resilient to the fronthaul delay, and hence the C-RAN and hybrid C-RAN are seen to achieve the best performance gain as compared to the conventional system.

In terms of efficiency, C-RAN outperforms the conventional system in all the considered cases due to the higher delivery rate per transmission offered by the interference management capabilities of the CU. Also, the efficiency of SW is comparable to that of SR, whereas the GBN protocol suffers due the additional transmissions of decoded packets that is due to the ``go-back" mechanism.

\begin{figure}[!t]
	\centering
	\includegraphics[width=\figwww]{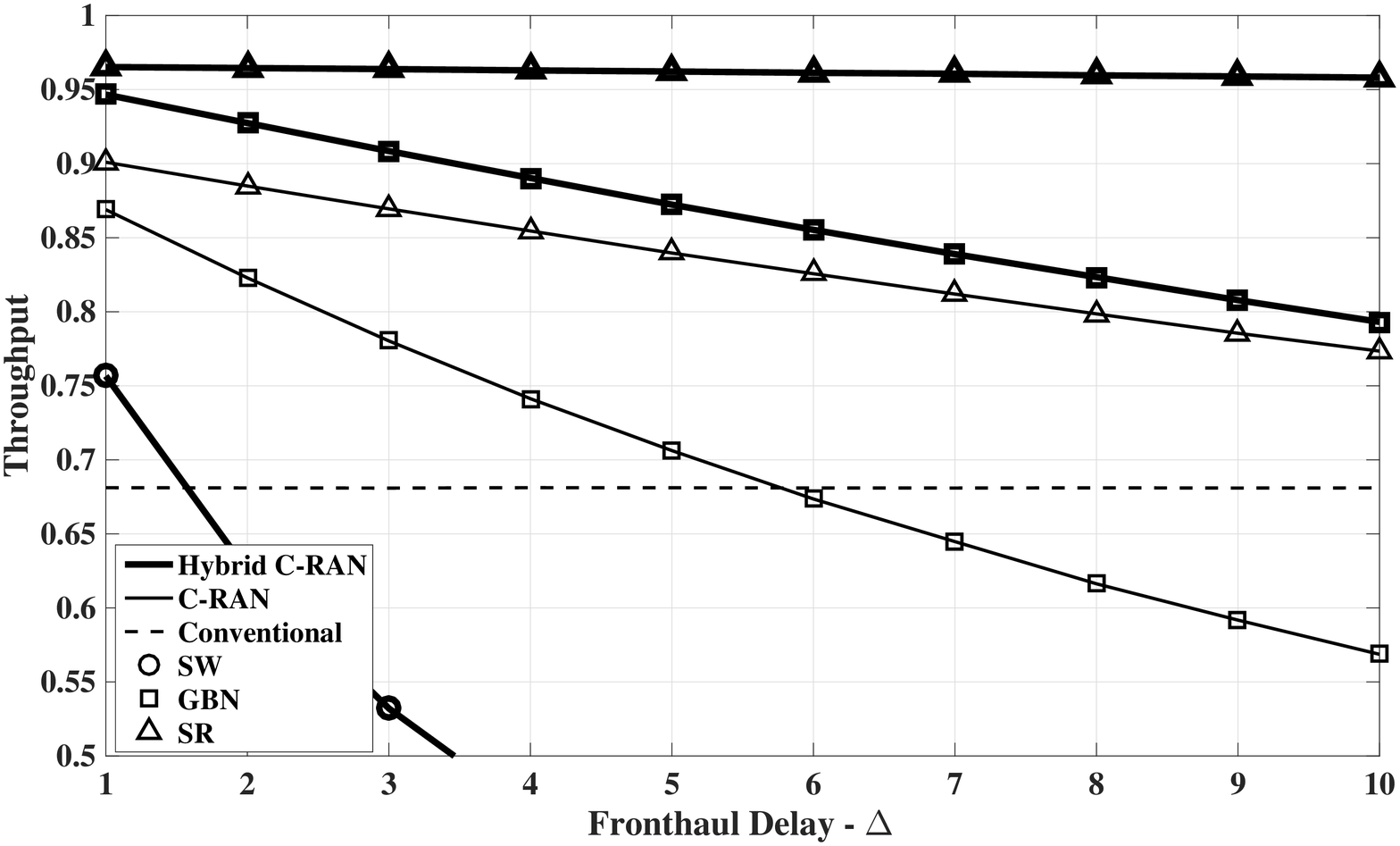}
	\caption{Throughput as a function of the fronthaul latency ($SIR{=}10$ dB, $\gamma{=}10$ dB, $\rho{=}0.3$, $\alpha{=}0.5$, $B_{\rm max}{=}1$, $W{=}\Delta$, $P{=}30$ dB, $E{=}0.6$ and $B{=}6$). \vspace{-8mm}}
	\label{figdelta}
\end{figure}

Fig.~\ref{figdelta} depicts the throughput as a function of the fronthaul latency $\Delta$. The performance of both the C-RAN and hybrid C-RAN system degrades as $\Delta$ increases. In particular, the SW ARQ mode system presents the largest sensitivity to the latency, as the latter entails an idle period, of duration $\Delta$, in which the channel is not used. Nevertheless, C-RAN or hybrid C-RAN still yield significant throughput gains with GBN and SR as long as $\Delta$ is not too large. For instance, for SR, C-RAN is advantageous over the conventional system for $\Delta \leq 6$.

\section{Conclusions}\label{conclusion}
This letter presents a study on the impact of fronthaul latency on the performance of the standard ARQ protocols Stop and Wait (SW), Go-Back-N (GBN) and Selective Repeat (SR)
for C-RAN systems. The analysis accounts for channel and interference temporal correlations, and offers a performance comparison in terms of throughput and efficiency with respect to traditional cellular systems and a hybrid C-RAN architecture where BSs attempts decoding before forwarding the received signal to the ``cloud''. The delay
introduced by the fronthaul network is observed to significantly degrade the throughput achieved by all ARQ protocols,
although SR presents the highest resilience to fronthaul latency. Moreover, the hybrid C-RAN system, which can be seen as an instance of alternative functional splits currently under study for C-RAN (see, e.g., \cite{5g,5g1}), is demonstrated to provide an effective means to mitigate the throughput and efficiency degradation caused by fronthaul latency. Interesting open problems include the analysis of the impact of multi-user detection in the presence of multiple MTs.

\end{document}